\documentclass[aps,twocolumn,superscriptaddress]{revtex4-1}
\usepackage{amsmath}
\usepackage{amssymb}
\usepackage{amsfonts}
\usepackage[overload]{empheq}
\usepackage{graphicx} 
\graphicspath{ {./Figs/} }
\usepackage[margin=0.75in]{geometry}
\usepackage[utf8]{inputenc}
\usepackage[english]{babel}
\usepackage[colorinlistoftodos]{todonotes}
\usepackage{mathtools}
\usepackage{float}
\usepackage{array,multirow}
\usepackage{bm}
\usepackage{cases}
\usepackage[position=top]{subfig}
\newcommand{\meff}{M_{\text{eff}}}
\newcommand{\tdl}{\tau_{\text{DL}}}

\newcommand{\xidl}{\xi_{\text{DL}}}
\newcommand{\xifl}{\xi_{\text{FL}}}
\newcommand{\xifmr}{\xi_{\text{FMR}}}

\newcommand{\tfm}{t_\text{F}}
\newcommand{\tnm}{t_\text{N}}

\newcommand{\Ramr}{R_\text{AMR}}

\begin{document}

\title{Resolving Discrepancies in Spin-Torque Ferromagnetic Resonance Measurements: Lineshape vs.\ Linewidth Analyses }
\author{Saba Karimeddiny}
    \email[Correspondence email address: ]{sk2992@cornell.edu}
    \affiliation{Cornell University, Ithaca, NY 14850, USA}
\author{Daniel C. Ralph}
    \affiliation{Cornell University, Ithaca, NY 14850, USA}
    \affiliation{Kavli Institute at Cornell, Ithaca, NY 14853, USA}

\date{\today} 

\begin{abstract}
When spin-orbit torques are measured using spin-torque ferromagnetic resonance (ST-FMR), two alternative ways of analyzing the results to extract the torque efficiencies -- lineshape analysis and analysis of the change in linewidth versus DC current -- often give inconsistent results.  We identify a source for these inconsistencies. We show that fits of ST-FMR data to the standard analysis framework leave significant residuals that we identify as due to (i) current-induced excitations of a small volume of magnetic material with magnetic damping much larger than the bulk of the magnetic layer, that we speculate is associated with the heavy-metal/magnet interface and (ii) oscillations of the sample magnetization at the modulation frequency due to heating. The dependence of the residual signals on DC current can interfere with an accurate extraction of spin-torque efficiencies by the linewidth method.  We show that the discrepancies between the two types of analysis can be largely eliminated by extrapolating the window of magnetic fields used in the linewidth fits to small values so as to minimize the influence of the residual signals. 
\end{abstract}

\maketitle

\section{Introduction}
Magnetic manipulation by spin-orbit torques (SOTs) \cite{Miron2011, Liu2012} is  a promising candidate mechanism for next-generation magnetic memory technologies \cite{Wang2013,Oboril2015}.  Accurate, consistent measurements of the efficiencies of SOTs are essential for fully understanding the microscopic origin of SOTs and for the development of SOT memory technologies; yet despite the large body of work that SOTs have amassed, quantitative discrepancies between different measurement techniques persist.  Here we will focus on inconsistencies that result from two approaches for analyzing spin-torque ferromagnetic resonance (ST-FMR) measurements, the most common experimental method used to quantify SOTs in heterostructures that have in-plane magnetic anisotropy \cite{Liu2011,Liu2012,Mellnik2014,Harder2011}.  These two approaches are lineshape (LS) analysis in which the in-plane and out-of-plane spin-orbit torques are determined from the amplitudes of the symmetric and antisymmetric resonance components \cite{Liu2011,Liu2012,Pai2015}, and linewidth (LW) analysis in which the damping-like in-plane component is determined from the dependence of the resonant linewidth on DC current \cite{Ando2008,Liu2011, Kasai2014, Nan2015}. These approaches often give results for the damping-like spin-orbit torque that differ by large factors (e.g., in the data we present below a difference by more than a factor of 3).

Here, we investigate the cause of the discrepancy by performing ST-FMR measurements on Pt/Permalloy and Pt/CoFeB samples, comparing data acquired using amplitude modulation of the microwave current (the most-common approach) and using frequency modulation, and by analyzing carefully the residuals in fits to the standard ST-FMR analysis for both modulation schemes.  We identify that the discrepancies between the lineshape and linewidth analyses arise primarily from current-induced excitation of a small volume of magnetic material with magnetic damping much greater than the majority of the magnetic film. This produces a low-amplitude background ST-FMR resonance with a broad tail, that can interfere with accurate determination of DC-current-induced changes in the linewidth of the primary resonance. We demonstrate that discrepancies between the lineshape and linewidth analyses can be reduced by employing frequency modulation rather than amplitude modulation of the microwave current, and by extrapolating the range of magnetic field used in the linewidth fits to small values so as to minimize the disruption from the low-amplitude, large-linewidth artifact signal.   

\section{Background on the ST-FMR technique}
During ST-FMR, a microwave current is applied in-plane into a heavy metal/ferromagnet bilayer, so that current-induced SOTs and {\O}rsted fields induce ferromagnetic precession. Magnetic precession causes resistance variations in the device due to anisotropic magnetoresistance. Mixing between the microwave current and resistance oscillations then produces a DC voltage that is measured. The signal-to-noise ratio (SNR) is vastly improved if an endogenous parameter of the technique is modulated and the voltage is measured with a lock-in amplifier. There are a number of choices for the modulated parameter: amplitude modulation (AM) \cite{Liu2011,Liu2012}, magnetic field modulation \cite{Goncalves2013},  frequency modulation (FM), or phase modulation (the latter two of which are equivalent).  Save a few works that have employed magnetic field modulation \cite{Goncalves2013, Skowronski2015, Safranski2016, Safranski2018, Chen2016, Xu2020}, almost all experiments featuring ST-FMR employ AM because it is the simplest parameter to modulate -- it does not complicate the experimental apparatus nor the fitted model. 

Typically, one assumes that a macrospin approximation is appropriate for describing the current-induced magnetic dynamics for experiments performed at suffficiently large microwave frequencies and magnetic fields, in which case the results of ST-FMR are modeled by the Landau-Lifshitz-Gilbert-Slonczewski (LLGS) equation
\begin{align}
    \dot{\mathbf{m}} = -\gamma \mathbf{m} \times \mathbf{B} 
                       + \alpha \mathbf{m} \times \dot{\mathbf{m}}
                       + \boldsymbol{\tau}
\end{align}
where $\mathbf{m}$ is the magnetic moment, $\gamma = 2\mu_B/\hbar$ is the gyromagnetic ratio, $\mathbf{B}$ is the external field, $\alpha$ is the Gilbert damping constant, and $\boldsymbol{\tau} = \boldsymbol{\tdl} + \boldsymbol{\tau_\text{z}}$ describes the torque present in our system. The torques produced by a polycrystalline thin film must obey the Rashba symmetry \cite{MacNeill2017}. For a film spanning the $X-Y$ plane, with current flowing along the $X$-direction we have
\begin{align}
    \boldsymbol{\tau_\text{DL}} &= \tau^0_\text{DL} (\mathbf{m} \times (\mathbf{m} \times Y)) 
                   =\frac{\xi_\text{DL}\mu_B}{e M_s \tfm}J_e\cos\phi_0 \label{tdl}\\
\begin{split}
    \boldsymbol{\tau_\text{z}} &= \tau^0_\text{z} (\mathbf{m} \times Y) \\
                  &= \left[\frac{\xi_\text{FL} \mu_B}{e M_s \tfm} + \frac{\mu_0 \tnm}{2}\right]J_e\cos\phi_0 \label{tfl}.
\end{split}
\end{align}
$\xi_\text{DL(FL)}$ is the damping-like (field-like) SOT efficiency, $\mu_B$ is the Bohr magneton, $M_s$ is the saturation magnetization of the ferromagnet, $t_\text{F(N)}$ is the thickness of the ferromagnet (normal metal) layer, $\mu_0$ is the vacuum permeability, $J_e$ is the electric current density flowing through the heavy metal, and $\phi_0$ is the angle between the direction of applied field and current flow ($X$-direction). Solutions to the LLGS equation for a sample with in-plane magnetic anisotropy predict that resonant ferromagnetic precession will occur when the Kittel equation, $\omega_0 = \gamma\sqrt{B(B+\mu_0\meff)}$ \cite{Kittel1948}, is satisfied. Here, $\mu_0\meff = \mu_0 M_s - 2K_\perp / M_s$ accounts for shape anisotropy minus any out-of-plane anisotropy. The total resonance lineshape will have contributions from symmetric ($S$) and antisymmetric ($A$) Lorentzians \cite{Harder2011,Liu2011}, which we define as
\begin{align}
   S = \frac{\Delta^2}{(B-B_0)^2 + \Delta^2} \\ 
   A = \frac{\Delta(B-B_0)}{(B-B_0)^2 + \Delta^2}
\end{align}
where $B_0$ is the resonance field, and $\Delta$ is the half-width-at-half-maximum linewidth related to the Gilbert damping by $\Delta =\alpha\omega/\gamma$. The DC mixing signal is a weighted sum of these two lineshapes with coefficients $V_S$ and $V_A$ determined by the the torques and material parameters in our system \cite{Liu2011, Harder2011,Karimeddiny2020}
\begin{align}\label{AMRes}
    V_\text{mix} = V_S S + V_A A + C
\end{align}
with
\begin{align}\label{VSVA}
\begin{split}
    V_S &= \frac{I_\text{rf}}{2\alpha\omega^+}R_\text{AMR}\tau^0_\text{DL}\sin2\phi_0\cos\phi_0 \\ 
    V_A &= \frac{I_\text{rf}}{2\alpha\omega^+}R_\text{AMR}\frac{\omega_2}{\omega}\tau^0_\text{z}\sin2\phi_0\cos\phi_0.
\end{split}
\end{align}
$I_\text{rf}$ is the total microwave current that flows through the bilayer, $\omega^+ = \gamma(2B_0 + \mu_0 \meff)$, $\Ramr$ is the amplitude of the anisotropic magnetoresistance of the whole bilayer, and $\omega_2 = \gamma(B_0 + \mu_0 \meff)$. $C$ is a constant voltage offset that is included to account for non-resonant signals. For samples with thick magnetic layers, there can also be a significant additional contribution to the symmetric resonance component from spin pumping/resonant heating and the inverse spin Hall effect \cite{Karimeddiny2020} that we will mention below. 

The experimental signal-to-noise ration (SNR) is significantly improved by modulating the microwave amplitude; this is captured by letting $V_\text{mix}(I_\text{rf}) \to V_\text{mix}(I_\text{rf}(1+ \mu\cos\omega_m t)) \approx V_\text{mix} + 2\mu V_\text{mix} \cos\omega_m t$ where $\mu \in \left[0,1\right]$ is the AM-depth.  A lock-in amplifier demodulates the total signal by mixing with a $\cos\omega_m t$ reference and applying a low-pass filter. The AM signal is therefore simply: 2$\mu V_\text{mix}$. 

If, alternatively, frequency modulation is used instead of amplitude modulation, the expected FM signal can be derived in a similar manner. We let $V_\text{mix}(\omega) \to V_\text{mix}(\omega + \delta \omega \cos\omega_m t)$ where $\delta \omega \ll \omega$; this admits the simple expansion near the microwave carrier frequency, $\omega_c$
\begin{align}\label{FMExp}
\begin{split}
    V^\text{FM}_\text{mix}(\omega) &= V_\text{mix}(\omega + \delta \omega \cos\omega_m t) \implies \\
    V^\text{FM}_\text{mix}(\omega) &\approx V_\text{mix}(\omega_c) 
    + \frac{d V_\text{mix}}{d \omega}\big\rvert_{\omega=\omega_c}\delta \omega \cos\omega_m t.
\end{split}
\end{align}
$V^\text{FM}_\text{mix}$ is again demodulated by a lock-in amplifier, which leaves us with only $\frac{d V_\text{mix}}{d \omega}\big\rvert_{\omega=\omega_c}\delta \omega$. Therefore, the ratio of the detected mixing signal to the amplitude of the frequency modulation is
\begin{align}\label{FMRes}
\begin{split}
    V^\text{FM}_\text{mix}/\delta \omega &=\frac{\partial V_\text{mix}}{\partial \omega}\big\rvert_{\omega=\omega_c}\\
     &= \frac{\partial V_S}{\partial \omega} S + \frac{\partial V_A}{\partial \omega} A \\
     &+ \frac{1}{\omega_c}\left[2V_S A^2 + V_A\left(2A^3/S - A\right) \right] \\
     &+ \frac{\omega_c}{\meff\gamma^2\Delta}\left[2V_S S A + V_A \left(A^2 - S^2\right)\right] + C.
\end{split}
\end{align}
Here we have used that $\partial_\omega S = 2S\left[(1 - S)\partial_\omega\Delta + A\partial_\omega B_0\right]/\Delta$ and $\partial_\omega A = \left[A(1 - 2S)\partial_\omega\Delta - (S+2A^2)\partial_\omega B_0\right]/\Delta$.
Equation (\ref{FMRes}) is nearly identical to a previously derived result where the magnetic field was modulated \cite{Goncalves2013}. Compared to the AM result, Eq.~(\ref{AMRes}), the FM result has two additional fit parameters, $dV_{S}/d\omega$ and $dV_{A}/d\omega$ to account for possible frequency dependence of microwave transmission through the measuring circuit to the device.   

\subsection{Lineshape (LS) Analysis}
After measuring with either amplitude or frequency modulation and then fitting the ST-FMR resonance to determine $V_S$ and $V_A$ using either Eq.~(\ref{AMRes}) or Eq.~(\ref{FMRes}), the torque efficiencies may be determined directly from Eq.~(\ref{VSVA}) if $I_\text{rf}$ is well-calibrated, since the other parameters in Eq.~(\ref{VSVA}) are independently-measurable.  However, since it is often challenging to determine accurately the value of $I_\text{rf}$ within the sample, we generally prefer to determine the torque efficiencies by taking appropriate ratios of $V_S$ and $V_A$ \cite{Pai2015}.  We first calculate an intermediate quantity, $\xifmr$, defined as
\begin{align}\label{xifmr}
    \xifmr &= \frac{V_S}{V_A}\frac{e \mu_0 M_s \tnm \tfm}{\hbar}\sqrt{1 +\frac{\mu_0 \meff}{B_0}}.
\end{align}
\noindent
By using Eqs.~(\ref{tdl}), (\ref{tfl}), \& (\ref{xifmr}), as long as the torque efficiencies are independent of the ferromagnetic-layer thickness in the range of thickness we analyze, then $\xifmr$ can be related to the damping-like and field-like torque efficiencies as   \cite{Pai2015}
\begin{align}\label{xidl}
    \frac{1}{\xifmr} &= \frac{1}{\xidl}\left(1 +\frac{\hbar}{e}\frac{\xifl}{\mu_0 M_s \tnm \tfm}\right).
\end{align}
Therefore, by taking results form a series of samples with different ferromagnet thicknesses, $\tfm$, we can then determine $\xidl$ and $\xifl$ from a linear fit of $1/\xifmr$ versus $1/t_F$.

\subsection{Linewidth (LW) Analysis: Change of Linewidth Versus DC Current}
In DC-biased ST-FMR, a DC current is applied parallel to the microwave current, such that the damping-like torque from the DC current rescales the effective Gilbert damping of the magnetic layer and causes the resonance linewidth to change linearly as a function of $I_\text{DC}$.   The damping-like SOT efficiency can be calculated from DC-current linewidth modulation as  \cite{Liu2011, Nan2015}
\begin{align}\label{xidlDC}
    \xidl &= \frac{e M_s \omega^+ \tfm}{\hbar \omega_c \sin\phi_0}\frac{W\tnm}{x}\frac{d \Delta}{d I_\text{DC}}
\end{align}
where $W$ is the width of the current-carrying channel and $x$ is the fraction of the total DC current that flows through the HM.

\section{Measurements}
All our samples are grown using DC-magnetron sputtering (in a system with base pressure $< 4\times10^{-8}$ torr) onto a surface-passivated high-resistivity Si wafer ($\rho >$  20,000 $\Omega$cm). Each sample is grown in an independent deposition. Samples shown in the main text have the stacking order: Substrate/Ta(1 nm)/Pt(6 nm)/Py($\tfm$)/Al(1 nm), with the magnetic layer being Permalloy (Py = Ni$_{81}$Fe$_{19}$). The Ta is used as a seed layer to promote smooth growth and the Al is oxidized upon exposure to air an is used as a capping layer to prevent oxidation of the Py. The Pt ($\rho$ = 20.4 $\mu\Omega$cm) and Py ($\rho$ = 25 $\mu\Omega$cm) are far more conductive than the Ta or oxidized Al so we assume all of the current flows through just the Pt and Py layers. Analogous results for which the Py is substituted with Co$_{40}$Fe$_{40}$B$_{20}$ can be found in the supplemental information.

After growth, we pattern the samples into rectangular bars of varying dimension using photolithography and Ar ion milling. The devices have dimensions: 40 $\mu$m $\times$ 80 $\mu$m, 20 $\mu$m $\times$ 60 $\mu$m, or 20 $\mu$m $\times$ 80 $\mu$m. All measurements shown in this work are taken on 20 $\mu$m $\times$ 80 $\mu$m devices, and the quantitative conclusions do not depend on the device geometry. We attach Ti(3 nm)/Pt(75 nm) contacts to the devices by another step of photolithography, DC-magnetron sputtering, and lift-off.

\begin{figure}[h]
\includegraphics[width=\linewidth]{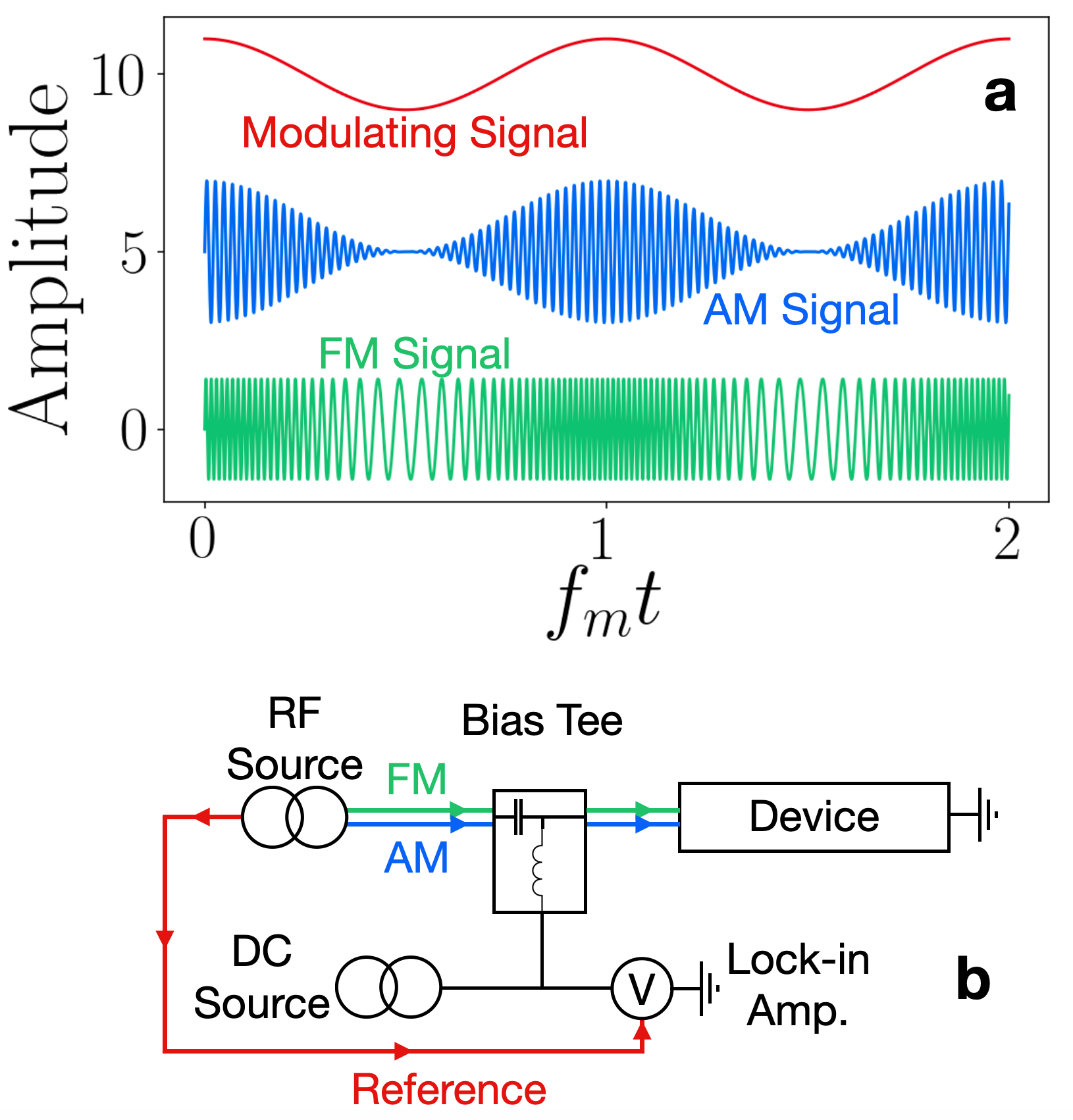}
\caption{The measurement setup used in this work. {\bf (a)} Schematic representations of the time dependence of microwave current that is injected into the device under test for amplitude-modulated and frequency-modulated experiments. Offsets are added and the scale of the frequency modulation is exaggerated for clarity. {\bf (b)} The circuit used in this measurement. The colors of the wires correspond to the colors of the signals in the top panel which the wire carries.}
\label{circuit}
\end{figure}

All data shown are measured on microwave-compatible Hall-bar structures that allow measurements of both longitudinal and transverse mixing voltages, as described in ref.~\cite{Karimeddiny2020}. Here, we will analyze only the longitudinal mixing voltages, as that is the usual ST-FMR measurement geometry. The devices are connected to the circuit shown in Fig.~\ref{circuit}(b). A RF source inputs a microwave current into the device through the AC port of a bias tee with either amplitude or frequency modulation (Fig.~\ref{circuit}(a)), while the magnitude of an external magnetic field is swept at a fixed angle $\phi_0$ through the Kittel resonance condition. The DC voltage along the longitudinal direction generated by mixing is detected with a lock-in amplifier that references the modulating signal. For the DC-biased measurements, an additional DC current is applied through the DC port of the bias tee to flow through the device in addition to the microwave current. All measurements are performed at room temperature.

\begin{figure*}[t]
\includegraphics[width=\linewidth]{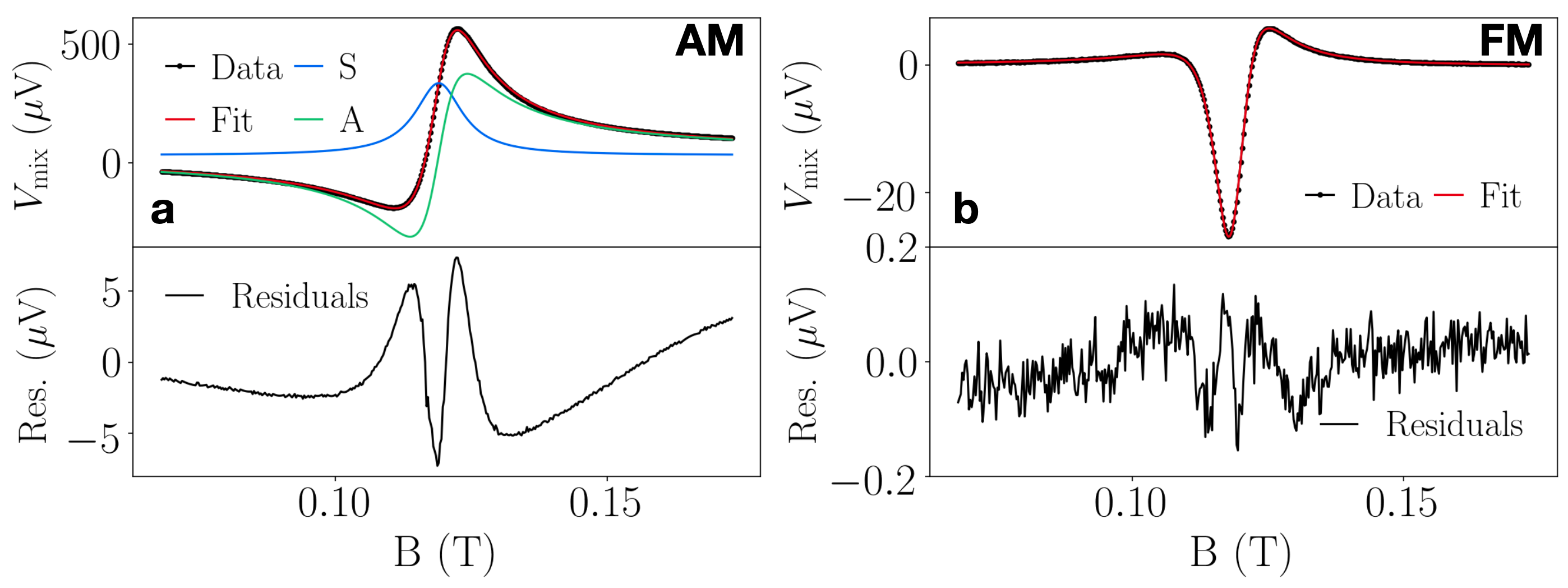}
\caption{Examples of measured resonances and fit residuals for a Pt(6 nm)/Py(5 nm) sample at 20 dBm, 10 GHz, $\phi_0 = 45^\circ$, $f_m=1.7$ kHz, $I_\text{DC}=0$. {\bf (a)} AM resonance taken at 100\% AM-depth with a fit to Eq.~(\ref{AMRes}). {\bf (b)} FM resonance taken $\Delta f = 16$ MHz with a fit to Eq.~(\ref{FMRes}).  The range of magnetic field shown here corresponds to [$B_0 - 15\Delta$, $B_0 + 15\Delta$] for the primary resonance. }
\label{res}
\end{figure*}

Measurements in the main text are performed at a 10 GHz carrier frequency ($f_c = \omega_c/2\pi$). The modulating signal for both AM and FM measurements is applied at 1.7 kHz modulation frequency ($f_m = \omega_m/2\pi$). The external magnetic field is applied at $\phi_0 = 45^\circ$ with respect to the direction of applied microwave current. The AM measurements are done with 100\% AM-depth as depicted in Fig.~\ref{circuit}(a) to maximize the measured signal. We find that reducing the AM-depth has no effect on the results shown (see supplemental information). The FM measurements are done with a frequency deviation ($\delta f = \delta\omega/2\pi$) of 16 MHz. Both the AM and FM are applied by the internal circuitry of the RF source, an Agilent 8257D. Both the 16 MHz frequency deviation and the 1.7 kHz modulation frequency are far smaller than the carrier frequency of 10 GHz, so that within either measurement mode the modulation has negligible effect on the microwave current over one precession cycle, a key assumption of the modeling.

\subsection{Results of Lineshape Analyses}
Examples of the longitudinal resonant mixing signals from a Pt(6 nm)/Py(5 nm) sample for both the AM and FM measurements are shown in Fig.~\ref{res}. The AM measurement (Fig.~\ref{res}(a)) is fit to Eq.~(\ref{AMRes}) with the five fit parameters $V_S, V_A, C, B_0, \Delta$ while the FM measurement (Fig.~\ref{res}(b)) is fit to Eq.~(\ref{FMRes}) including the two additional fit parameters $dV_{S}/d\omega, dV_{A}/d\omega$. The fit to the AM measurement looks good by eye, but the best fit nevertheless produces significant systematic residuals, which hints that the framework of conventional ST-FMR analysis (Eq.~(\ref{AMRes})) gives an incomplete description. To rule out spurious measurement artifacts, we have repeated the AM measurements on three independent ST-FMR apparatuses at Cornell and have also performed measurements on different sample stacks; all of these measurements show the same systematic residuals for the AM fits. In contrast, for the FM measurements the scale of the residuals after fitting to Eq.~(\ref{FMRes}) is significantly smaller relative to the full signal magnitude.  A more complete discussion of the fit quality, residuals, and statistical details can be found in the supplemental information.

For the AM fits in Fig.~\ref{res}(a), we see that the residuals have a lineshape near the resonance field that closely resembles a Lorentzian derivative, suggesting that an additional parameter in Eq.~(\ref{AMRes}) is varying at the modulation frequency and contributing to the homodyne mixing signal. Quantitative estimates suggest that a varying $\meff$ will contribute far more than other candidate sample parameters and that an $\meff$ oscillating at the AM frequency, presumably due to heating, can result in the residual lineshape we observe near the resonance field. That is, suppose  (in addition to the amplitude modulation of $I_{RF}$) that $\meff$ also varies periodically as $\meff \to \meff + \delta\meff\cos\omega_m t$; this, analogously to the frequency modulation, would allow the expansion
\begin{align}
\begin{split}
    &V_\text{mix}(\meff + \delta\meff\cos\omega_m t) \\
    &\approx V_\text{mix}(\meff) + \frac{\partial V_\text{mix}}{\partial \meff} \delta\meff \cos\omega_m t.
\end{split}
\end{align}
The total mixing signal will thus consist of the sum of two terms that vary periodically with the AM
\begin{align}
\begin{split}
   \left(2\mu V_\text{mix} + \frac{\partial V_\text{mix}}{\partial \meff} \delta\meff\right)  \cos\omega_m t
\end{split}
\end{align}
 and \textit{both} will be demodulated by the lock-in amplifier. 

A homodyne signal from an oscillating value of $\meff$ cannot by itself explain the full residual in the AM fits; in addition the AM residuals appear to contain an ordinary AM resonance lineshape (Eq.~(\ref{AMRes})) with a very large linewidth.
\begin{figure}[h!]
\includegraphics[width=\linewidth]{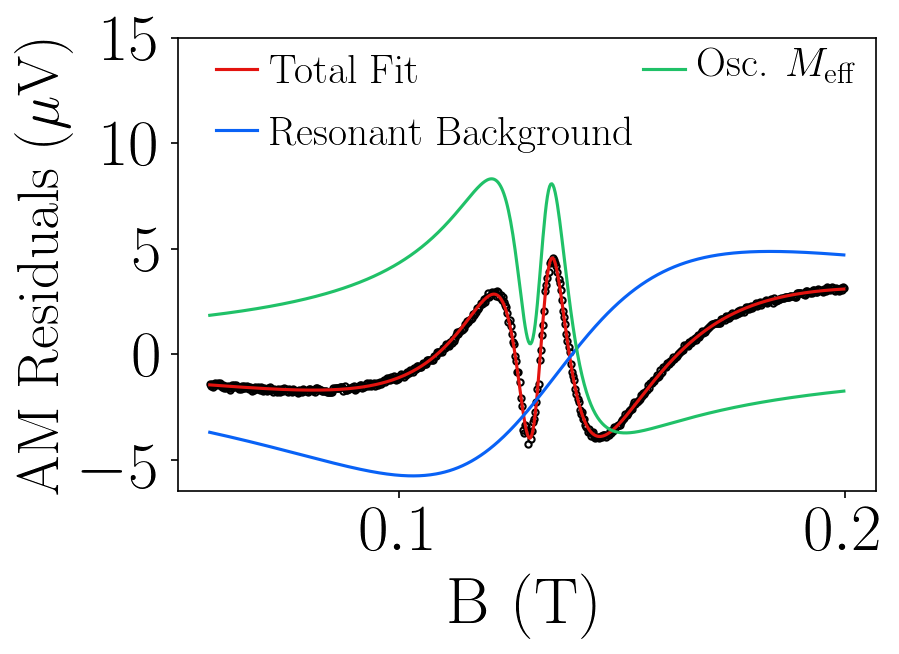}
\caption{{\bf (a)} Measured fit residuals for an AM measurement of a Py(3 nm) sample taken at zero DC bias, with a fit to the sum of contributions from a heating-induced oscillation of $M_\text{eff}$ and a large-$\Delta$ resonant background. The linewidth of the primary resonance for Py(3 nm) samples is greater than for Py(5 nm) samples, so the range of magnetic field shown here still corresponds to [$B_0 - 15\Delta$, $B_0 + 15\Delta$] for the primary resonance.}
\label{MeffT}
\end{figure}
In Fig.~\ref{MeffT} we show the fit residuals of an AM measurement taken on a Py(3 nm) sample (with no DC current bias).  We fit the residuals to the sum of a homodyne signal corresponding to an oscillating value of $\meff$ (green curve) and a large-$\Delta$ resonant background (Eq.~(\ref{AMRes})) (blue) with $\Delta_\text{large} = 41.6$ mT, much larger than the value $\Delta = 7.3$ mT for the primary resonance.  The sum of the two contributions (red curve) fits the residuals very well.  Based on direct measurements of $\meff$ versus temperature on the same device [$d(\mu_0\meff)/d T = 8 \times 10^{-4}$ T/$^\circ$C], the scale of the temperature oscillations needed to produce the oscillating-$\meff$ homodyne signal is approximately $1^\circ$C (see supplemental information). $V_S$ and $V_A$ for the large-$\Delta$ resonance for the data in Fig.~2 are 12 $\mu$V and 7 $\mu$V, while for the primary resonance $V_S$ = 311 $\mu$V and $V_A$ = 684 $\mu$V.

We have considered two options for the origin of the large-$\Delta$ resonance: a region of increased damping (a) near the sample edges or (b) near a magnetic interface.  If the origin were due to increased damping near the sample edges, we would expect the ratio of the amplitudes for the large-$\Delta$ and primary resonances to scale inversely with the sample width and to be approximately independent of the ferromagnetic-layer thickness.  Instead, we find that this ratio is insensitive to the sample width (a change of $< 10 \%$ in the symmetric and $< 4 \%$ in the antisymmetric component upon changing the sample width by a factor of 2), while it is sensitive to the ferromagnetic layer thickness (see supplemental information).  This suggests that the portion of the sample with increased damping is an interfacial region. Additional evidence for an origin associated with the heavy-metal/ferromagnet interface comes from the fact that the large-$\Delta$ linewith is very sensitive to applied DC current (see supplemental information), consistent with a very thin and/or low-moment region under the influence of the spin current generated by the heavy metal.  Our observations might be related to recent findings from the IBM group of interfacial regions in CoFeB/MgO/CoFeB magnetic tunnel junctions whose dynamics can become partially decoupled from the bulk of the magnetic films \cite{sun2017,safranski2019}. The two experiments differ, however, in that the IBM work deduced a difference in effective magnetic anisotropy (compared to the bulk of the magnetic film) for the interfacial layers at CoFeB/MgO interfaces, while in our devices the large-$\Delta$ resonance corresponds to an increased damping near a Pt/ferromagnet interface without a large difference in anisotropy. 

We suggest that there are two reasons why the fit residuals for the FM measurements are reduced compared to the AM measurements.  First, temperature oscillations at the modulation frequency will be smaller for the FM measurements because the magnitude of $I_\text{RF}$ will be approximately constant in time, so Ohmic heating caused by $I_\text{RF}$ will also be approximately constant rather than oscillating at the modulation frequency.  Temperature oscillations will not be eliminated completely however, since FM near the resonance will cause the energy absorbed by resonant heating of the magnetic layer (energy transfer associated with magnetic excitation by the current-induced torques) to oscillate at the modulation frequency.  We suggest that this resonant heating is likely the main cause of the small remaining systematic residuals near the resonance field in the fits to the FM data (Fig.~\ref{res}(b)).  Second, contributions from the large-$\Delta$ resonance to the FM measurements are reduced precisely because the linewidth is so broad, so this part of the signal is relatively insensitive to variations in applied frequency.

\begin{figure}[h!]
\includegraphics[width=\linewidth]{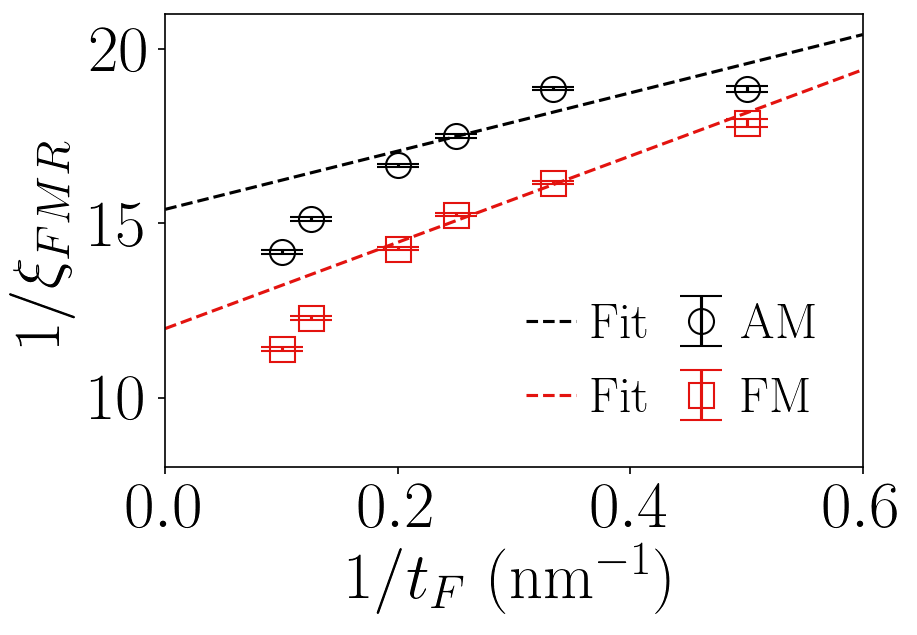}
\caption{Fits to Eq.~(\ref{xidl}) to determine the spin torque efficiencies $\xidl$ and $\xifl$ from the lineshape analyses.  The values of $\xifmr$ plotted are calculated according to Eq.~(\ref{xifmr}), corresponding to measurements of $V_S$ and $V_A$ done at 20 dBm, 10 GHz, $\phi_0 = 45^\circ$, $f_m=1.7$ kHz, and $I_\text{DC}=0$. The resonance fits are performed over the same window as the LW measurements: [$B_0 - 15\Delta$, $B_0 + 15\Delta$]. AM measurements are done with 100\% AM-depth and FM measurements with $\delta f = 16$ MHz.}
\label{pai}
\end{figure}

If one proceeds with the standard ST-FMR macrospin analysis -- Eqs.~(\ref{xifmr}), (\ref{xidl}) -- (ignoring the residuals for now) the resulting values of $1/\xifmr$ for both the AM and FM measurements are shown in Fig.~\ref{pai} for samples with ferromagnet layer thicknesses $t_F$ varying from 2 nm to 10 nm.  The samples with the thickest ferromagnet layers ($t_F \geq8$ nm) show deviations from a linear dependence of $1/\xifmr$ vs.\ $1/t_F$ that can be understood as due to the effect of an inverse spin Hall voltage resulting from spin pumping or resonant heating \cite{Pai2015, Karimeddiny2020}.  We therefore perform the linear fits only to the four samples with the thinnest F layers, extracting the values shown in Table \ref{ls_result}.
$\xidl$ and $\xifl$ are calculated from the y-intercept and slope of the  fits, respectively, following the prescription of Eq.~(\ref{xidl}). The FM and AM methods yield values for both $\xidl$ and $\xifl$ that differ by considerably more than the estimated statistical uncertainty in the results.  The difference in the values of $\xidl$ is about 30\%, while for $\xifl$ the FM result nearly double that of the AM. 

\begin{table}[h!]
    \centering
    \begin{tabular}{c|cc}
         & AM  & FM \\
            \hline
        $\xidl$  &  0.0650(4) & 0.0835(7)\\
        $\xifl$ & 0.0050(2) & 0.0094(2)
    \end{tabular}
    \caption{$\xidl$ and $\xifl$ that result from the linear fits shown in Fig. \ref{pai}.}
    \label{ls_result}
\end{table}

We suggest that the differences between these AM and FM LS results can be explained by the neglect of the residual terms.  If we take the values of $\xidl$ and $\xifl$ determined by the FM measurements and use them in fitting to the AM data, the result is a residual similar to that shown in Fig.\ \ref{MeffT} that can be fit just as well to a sum of a signal due to an oscillating value of $M_\text{eff}$ plus a large-$\Delta$ resonance (see supplemental information).  Fits to the AM data that include both the primary resonance and the two artifact contributions therefore possess near-degenerate fitting parameters, that can make determination of the spin-torque efficiencies imprecise.

\noindent

\subsection{Results of Linewidth Analyses}
\begin{figure}[h]
\includegraphics[width=\linewidth]{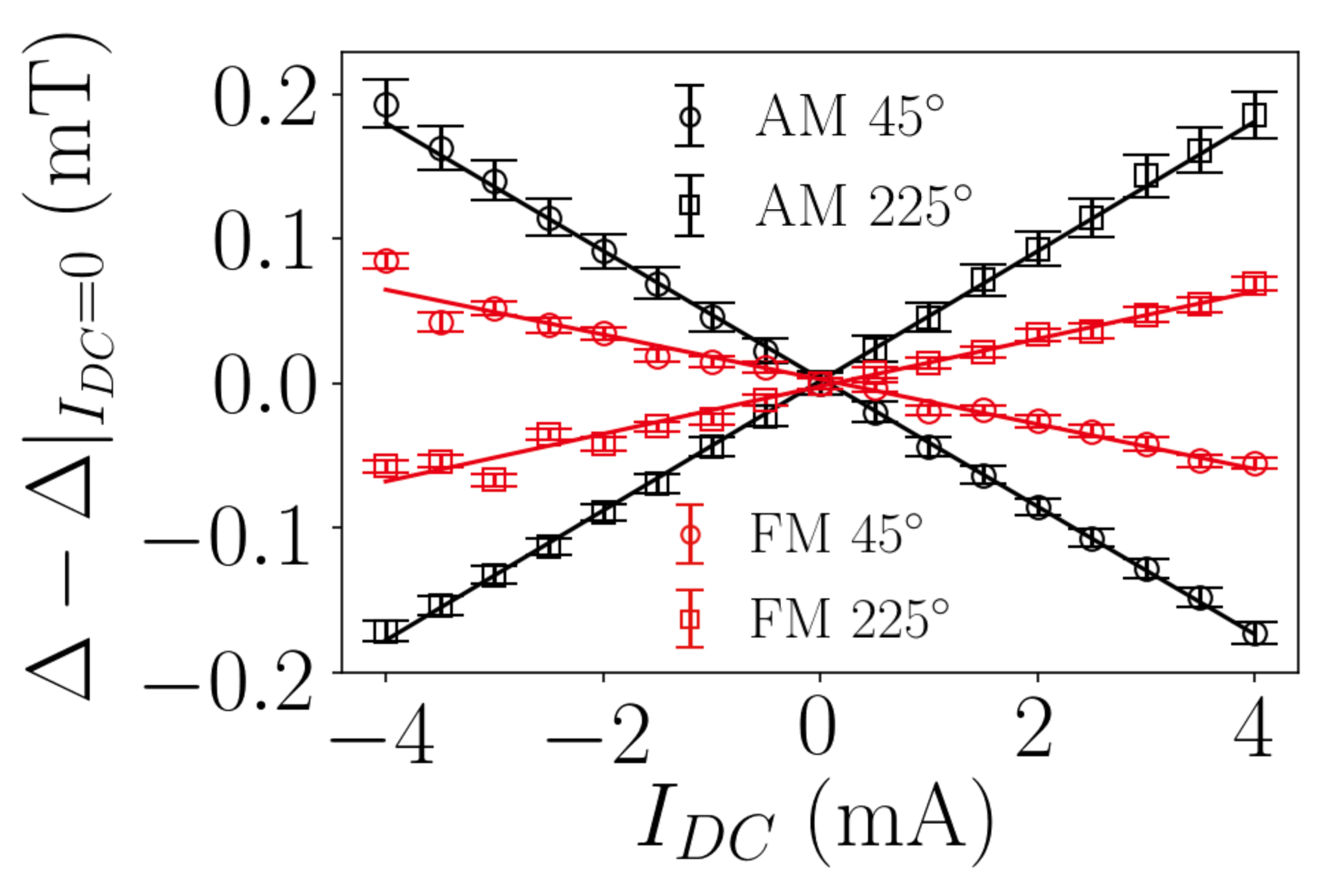}
\caption{Dependence of the resonance linewidth ($\Delta$) on $I_\text{DC}$ for a Pt(6 nm)/Py(5 nm) sample based on fits to Eq.~(\ref{AMRes}) for the AM measurements and Eq.~(\ref{FMRes}) for the FM measurements, for data collected  at 20 dBm, 10 GHz, $\phi_0 = 45^\circ/225^\circ$, $f_m=1.7$ kHz.  Linewidths are extracted using a fit window [$B_0 - 15\Delta$, $B_0 + 15\Delta$]. The zero-DC-current linewidths (5.27 mT for AM and 5.24 mT for FM) are subtracted. The solid lines are least-squares best fit lines to the data.}
\label{cross}
\end{figure}

The LW measurement proceeds identically to the LS measurement, but for the application of a DC current parallel to the microwave current. A full resonance lineshape (e.g., Fig.~\ref{res}) is collected for DC currents ranging from $-$4 mA to 4 mA and the resonances are fit to Eq.~(\ref{AMRes}) for the AM measurements or Eq.~(\ref{FMRes}) for the FM measurements to extract the linewidth, $\Delta$ (ignoring residuals for now). We find that the value of $\Delta$ that we get from the fits for the AM measurement depends strongly on whether and to what extent we include the tails of the resonance. Figure \ref{cross} shows the current dependence of linewidths for a Pt(6 nm)/Py(5 nm) sample extracted from fits over the field range [$B_0 - 15\Delta$, $B_0 + 15\Delta$] (with $\Delta$ adjusted for each sample corresponding to the linewidth of the primary resonance at zero DC current).  This is the largest fit window that is possible while consistently excluding artifacts associated with deviations from magnetic saturation at low field for all samples.    The zero-current value of $\Delta$ is subtracted from each of the plots in Fig.\ \ref{cross} to highlight the difference in the slopes of the best-fit lines. We apply Eq.~(\ref{xidlDC}) to the slopes of the best-fit lines and get the results for $\xidl$ shown in Table \ref{dc_result}.

\begin{figure}[h!]
\includegraphics[width=\linewidth]{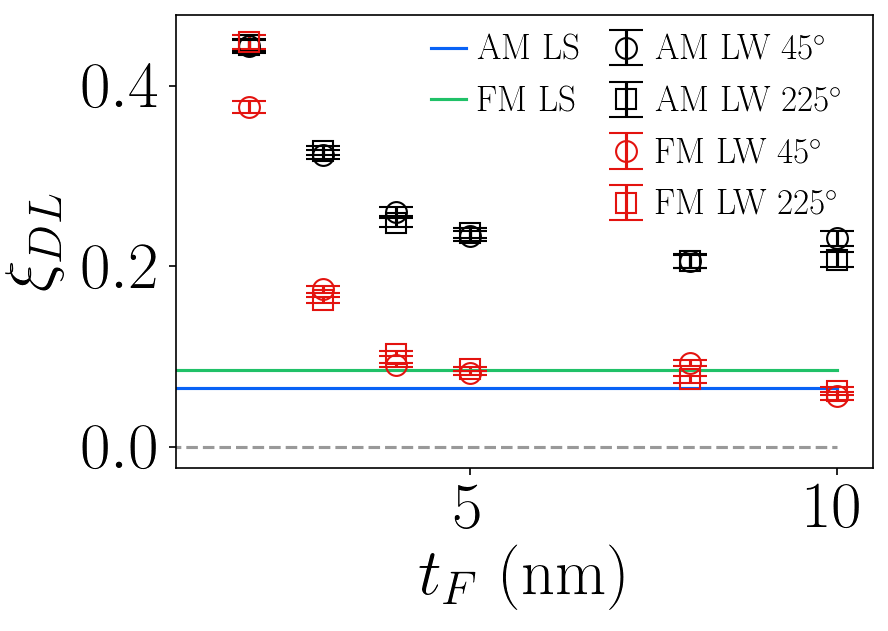}
\caption{Extracted values of the damping-like spin-torque efficiency $\xidl$ for samples with different ferromagnet layer thicknesses.  Symbols show the results of the AM and FM linewidth analyses using fits over the field range [$B_0 - 15\Delta$, $B_0 + 15\Delta$]. The green and blue lines are the results of the lineshape analyses for the thickness series shown in Fig.~\ref{pai}. }
\label{LWThick}
\end{figure}

For this sample, we see that the FM LW measurements agree with the FM LS result within the experimental uncertainties  (Table I), while the AM LW measurements differ by more than a factor of 3 from both the FM results and the AM LS measurements. Figure \ref{LWThick} compares the results of similar LW analyses for all of the Pt(6 nm)/Py($\tfm$) samples with different magnetic-layer thicknesses using the same fit window [$B_0 - 15\Delta$, $B_0 + 15\Delta$].
The AM LW measurements (black points) give far larger values for $\xidl$ compared to any of the other techniques.  The FM LW measurements are reasonably consistent with the FM LS value in the range $t_F =$ 4 - 10 nm (with small deviations for $t_F = 10$ nm possibly due to the neglect of an inverse spin Hall voltage generated by spin pumping or a spin Seebeck effect), but the FM LW measurements also differ increasingly from the the LS results for Py thicknesses below 4 nm.

\noindent

\begin{table}[h]
    \centering
    \begin{tabular}{c|cc}
        {\bf $\xidl$} & AM  & FM \\
            \hline
        45$^\circ$  &  0.234(5) & 0.082(2)\\
        225$^\circ$ & 0.237(5) & 0.087(2)
    \end{tabular}
    \caption{Table of $\xidl$ values for a Pt(6 nm)/Py(5 nm) sample using the LW method. The values are extracted from the slopes of the best fit lines in Fig.~\ref{cross} and Eq.~(\ref{xidlDC}).}
    \label{dc_result}
\end{table}

\begin{figure}[h]
\includegraphics[width=\linewidth]{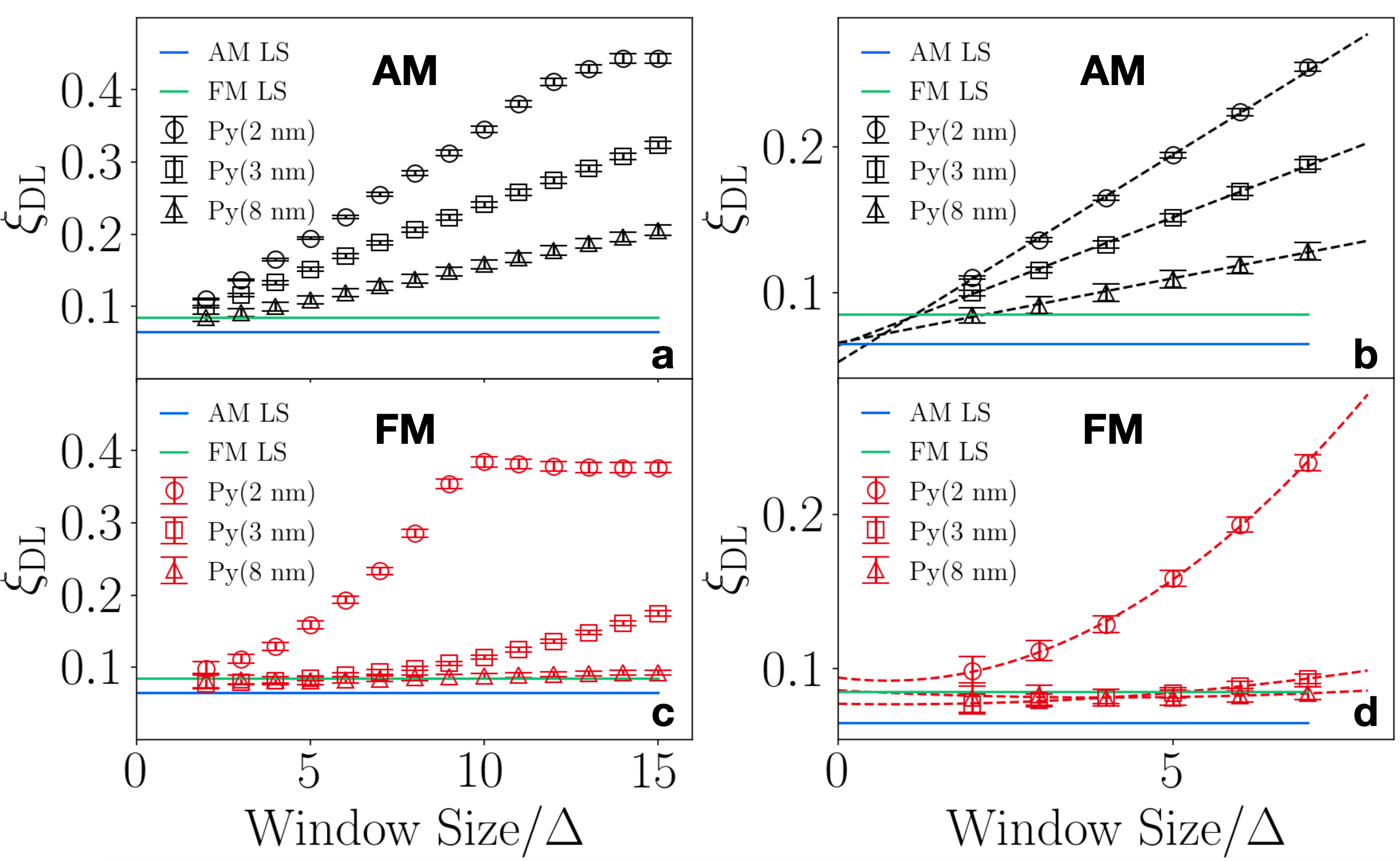}
\caption{The DL-SOT efficiency, $\xidl$, vs. the size of the fit window (normalized by the resonance linewidth, $\Delta$). All data in this figure is from the LW analysis method. {\bf (a)} The full range of fit windows with AM. {\bf (b)} A zoomed view of the AM data with best-fit lines superimposed. $\xidl$ is linear in the fit window and the y-intercept of the best-fit lines agrees well with the corresponding result of the AM LS analysis. {\bf (c)} The full range of fit windows with FM. {\bf (d)} A zoomed view of the FM data with best-fit second-degree polynomials superimposed. $\xidl$ is quadratic in the fit window and the y-intercept of the best-fit lines agrees well with the corresponding result of the FM LS analysis.}
\label{extrap}
\end{figure}

In Fig.\ \ref{extrap} we show the results of the same LW analysis using different sizes for the window of magnetic field included in the fits.  The panels on the left show the values of $\xidl$ extracted for window sizes from [$B_0 - 15\Delta$, $B_0 + 15\Delta$] to [$B_0 - 2\Delta$, $B_0 + 2\Delta$]. For both the AM and FM data sets, the extracted values of $\xidl$ decrease with decreasing window size.  We interpret this dependence as a clear indication that the LW analysis can be disrupted by the long tails of the residual terms that are not included as part of the standard linewidth analysis.  For a fixed value of fit-window size, the disruption is most severe for magnetic layers thinner than 4 nm because the linewidth of the primary resoance increases for thin layers, making the primary resonance more difficult to disentangle from the large-linewidth residual signal.  The right panels of Fig.\ \ref{extrap} show zoom-ins of the same LW results to better visualize the extrapolation of the measurements to zero linewidth.  We find that this extrapolation brings the results of both the AM LW and FM LW analyses into reasonable quantitative agreement with the  lineshape results.

We emphasize that the sensitive dependence on fit-window size shown by Fig.\ \ref{extrap} occurs despite the fact that the individual fits look quite good by eye for any choice of window size. The LW analyses are based on quite subtle changes in the resonance lineshape, e.g. about a 2\% change in linewidth over the full range of $I_\text{DC}$ for the FM measurements shown in Fig.~\ref{cross}.  Therefore, even small changes in $V_\text{mix}$ associated with current-dependent residuals can affect the LW analysis -- the small tails of the ST-FMR resonances can be substantially affected even if the overall magnitude of the residual signals near the resonance field is small. The large-$\Delta$ resonance in particular has a large affect on the LW analyses because its linewidth is strongly current dependent (see supplemental Fig.\ 13). We have tried fitting the AM resonances to a generalized Eq.\ \ref{AMRes} that includes the models for the residuals directly in the fit, but this is not able to provide improved quantitative results because of near-degenerate fit parameters.  We therefore recommend the procedure depicted in Fig.\ \ref{extrap} as the simplest approach to improving ST-FMR linewidth analyses -- performing the standard ST-FMR fits using a series of different fit-window sizes and then extrapolating to small windows to minimize the influence of the large-linewidth residuals.

\section{Conclusions}
We have identified a cause of inconsistencies between measurements of spin-orbit torque determined via lineshape and linewidth analyses of ST-FMR data -- that the standard model for analyzing ST-FMR data does not fully account for all of the magnetic dynamics that can affect the measurements. The standard analysis leaves residuals that we identify as due to (i) current-induced excitation of a magnetic mode with larger damping than the bulk of the magnetic layer and also (ii) temperature oscillations ($\approx 1$ $^\circ$C) associated with the modulation schemes employed for lock-in amplifier measurements.  The residuals are not large, with amplitudes of order 1\% of the primary resonance, but nevertheless they can affect the current dependence of the resonance tails sufficiently to disrupt an extraction of the anti-damping spin-orbit torque efficiency based on the current dependence of the ST-FMR linewidth. The influence of the large-linewidth residuals can be minimized by performing the standard lineshape analysis using different choices for the range of magnetic field values used to fit the ST-FMR resonances, and then extrapolating to zero fit window.  We recommend this procedure for all future uses of the LW analysis.  The effect of the residuals can also be reduced by performing ST-FMR using frequency modulation rather than amplitude modulation, but frequency modulation alone does not cure inconsistencies between the lineshape and linewidth results for our thinnest magnetic layers without extrapolation of the fit window to small values.

It remains an interesting open question what is the microscopic origin of the large-linewidth mode that contributes to the residual signal.  Based on the scaling of signal amplitudes with the widths and thicknesses of our samples, we identify this mode with the heavy-metal/magnet interface rather than as due to increased damping at the lateral edges of our magnetic layers. It is possible that this mode is due to an interface magnon, magnetic impurities caused by intermixing near the interface, a magnetic proximity layer within the platinum, or coupled dynamics involving two or all three of these effects. We plan future experiments to begin to resolve this question by making samples with different heavy metals and performing measurements as a function of temperature, to vary the importance of magnetic proximity effects.

\section{Acknowledgments}
This research was funded by the US Dept.\ of Energy (DE-SC0017671).  The work was performed in part at the Cornell NanoScale Facility, a member of the National Nanotechnology Coordinated Infrastructure (NNCI), which is supported by the National Science Foundation (Grant NNCI-2025233), and in part in the Cornell Center for Materials Research Shared Facilities which are supported through the NSF MRSEC program (DMR-1719875). 


\bibliographystyle{apsrev4-1}
\bibliography{bibl}


\pagebreak

\onecolumngrid
\begin{center}
  \textbf{\large Supplementary Information: \\ Resolving Discrepancies in Spin-Torque Ferromagnetic Resonance Measurements: Lineshape vs.\ Linewidth Analyses}\\[.2cm]
  Saba Karimeddiny,$^{1,*}$ and Daniel C. Ralph$^{1,2}$\\[.1cm]
  {\itshape ${}^1$Cornell University, Ithaca, NY 14850, USA\\
  ${}^2$Kavli Institute at Cornell, Ithaca, NY 14853, USA\\}
  ${}^*$Correspondence email address: sk2992@cornell.edu\\
(Dated: \today)\\[1cm]
\end{center}
\onecolumngrid

\setcounter{equation}{0}
\setcounter{figure}{0}
\setcounter{table}{0}
\setcounter{page}{1}
\setcounter{section}{0}
\renewcommand{\theequation}{S\arabic{equation}}
\renewcommand{\thefigure}{S\arabic{figure}}


\section{Properties of the Permalloy layer as a function of thickness}
\begin{figure}[h]
\includegraphics[width=0.7\linewidth]{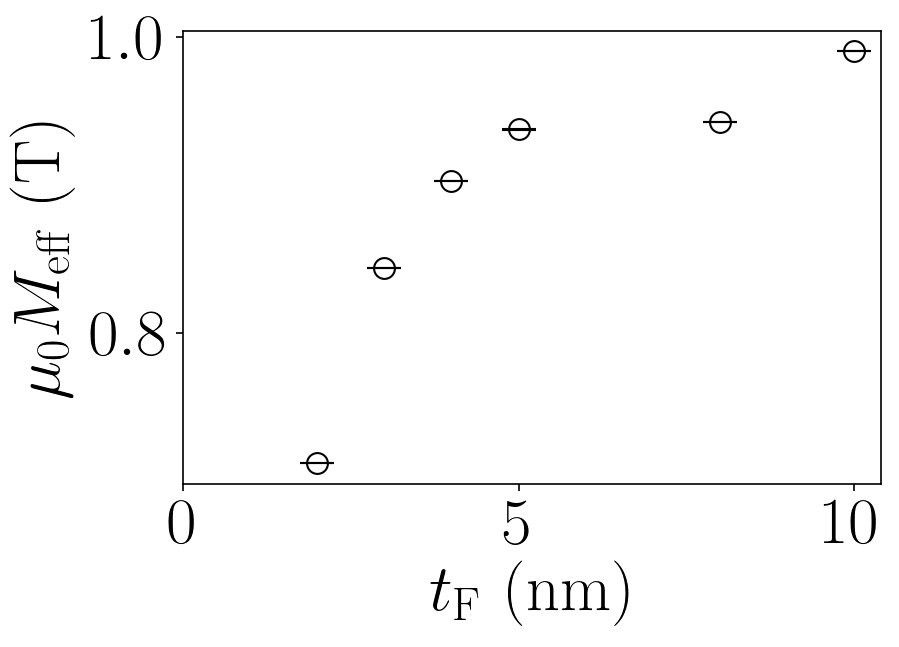}
\caption{$\mu_0 M_\text{eff}$ vs. Py thickness in nm. The thinner magnetic layers have larger out-of-plane anisotropy, which reduces the value of $M_\text{eff}$. The data increases to saturate at around 1 T ($\approx M_s$).}
\label{meff}
\end{figure}
\begin{figure}[h]
\includegraphics[width=0.7\linewidth]{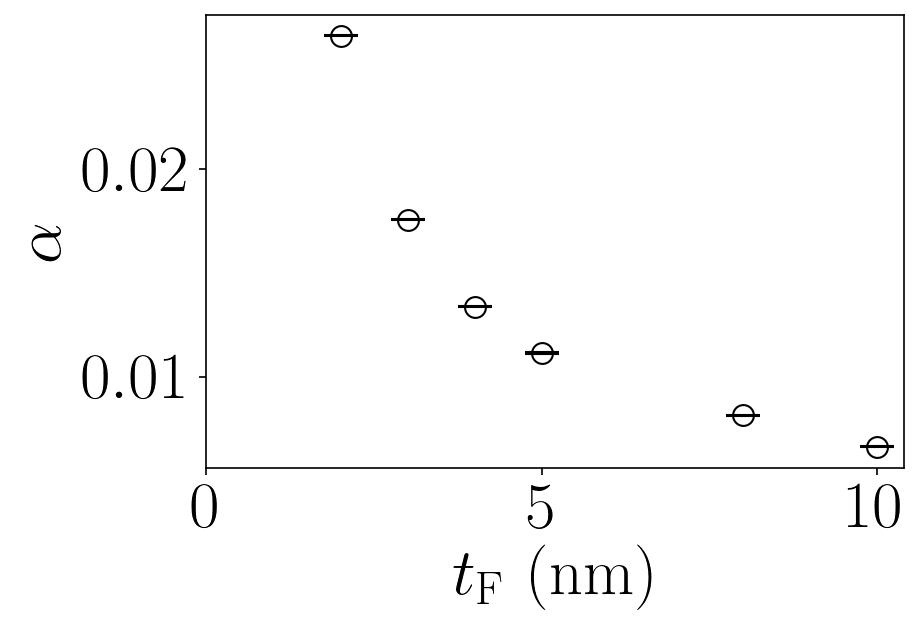}
\caption{Zero-bias Gilbert damping paramter, $\alpha$ vs. Py thickness in nm.}
\label{alpha}
\end{figure}
\begin{figure}[h]
\includegraphics[width=0.7\linewidth]{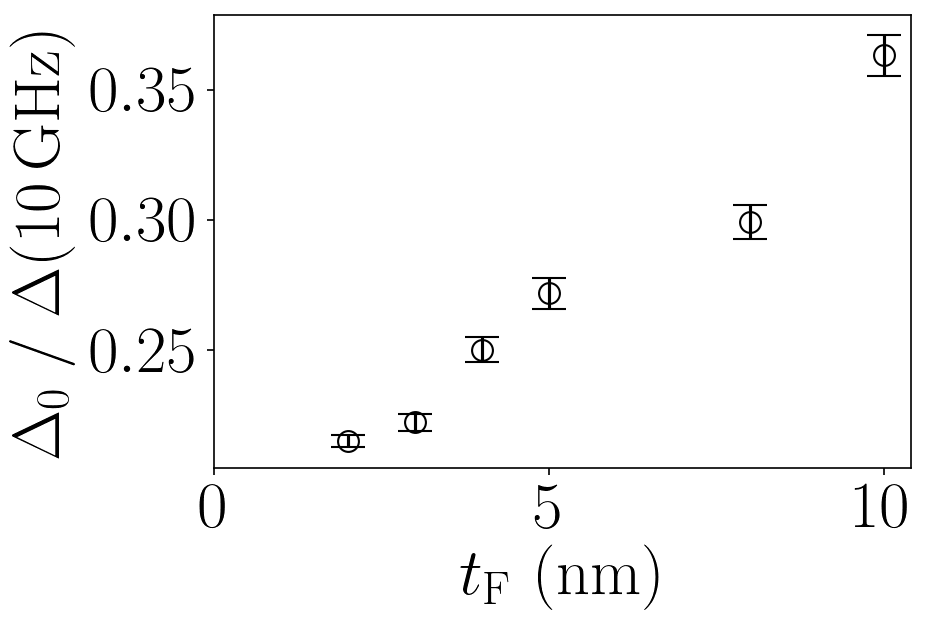}
\caption{Zero-bias inhomogenous resonance linewidth, $\Delta_0$ (normalized by the 10 GHz zero-bias resonance linewidth) vs. Py thickness in nm.}
\label{delta0}
\end{figure}
\begin{figure}[h]
\includegraphics[width=0.7\linewidth]{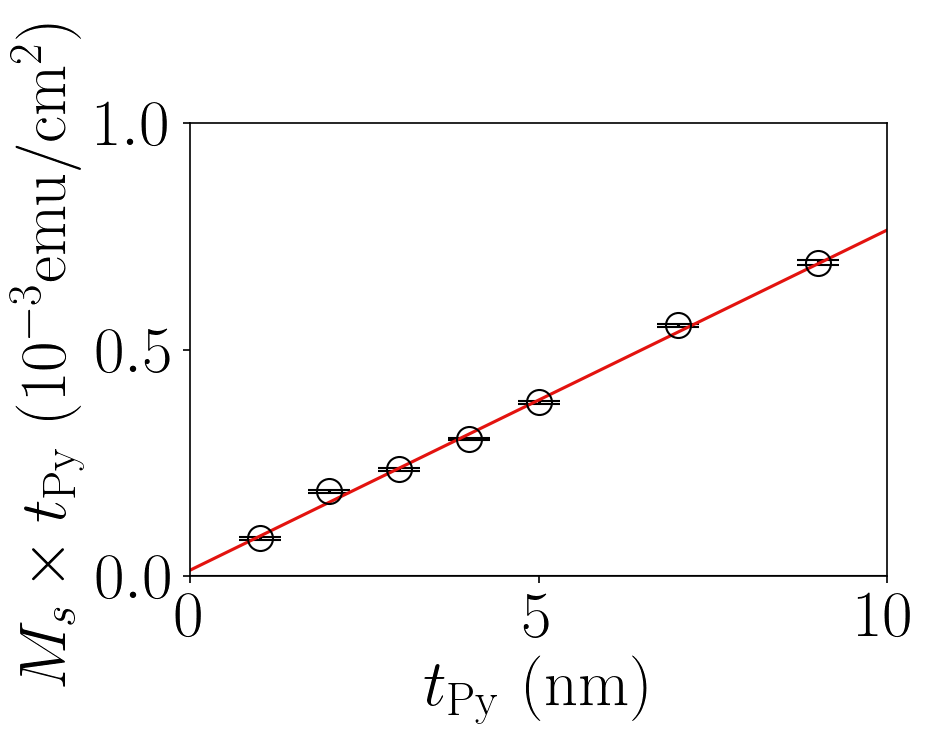}
\caption{Saturation magnetization per unit Py area vs. Py thickness in nm measured via vibrating sample magnetometry (VSM). The slope gives an accurate measurement of $M_s=0.95$ T and the x-intercept is within fit error of 0, indicating a negligible magnetic dead layer thickness.}
\label{vsm}
\end{figure}

\clearpage

\section{Fit of the AM residual determined using the torque efficiencies from the FM measurement}
In this section, we show an AM resonance where the fit has been performed with a fixed torque ratio (S/A) determined by the fit to the FM data. The residuals are still fit well by the sum of a function describing an oscillating $\meff$ and large-$\Delta$ resonance.
\begin{figure}[h!]
\includegraphics[width=0.7\linewidth]{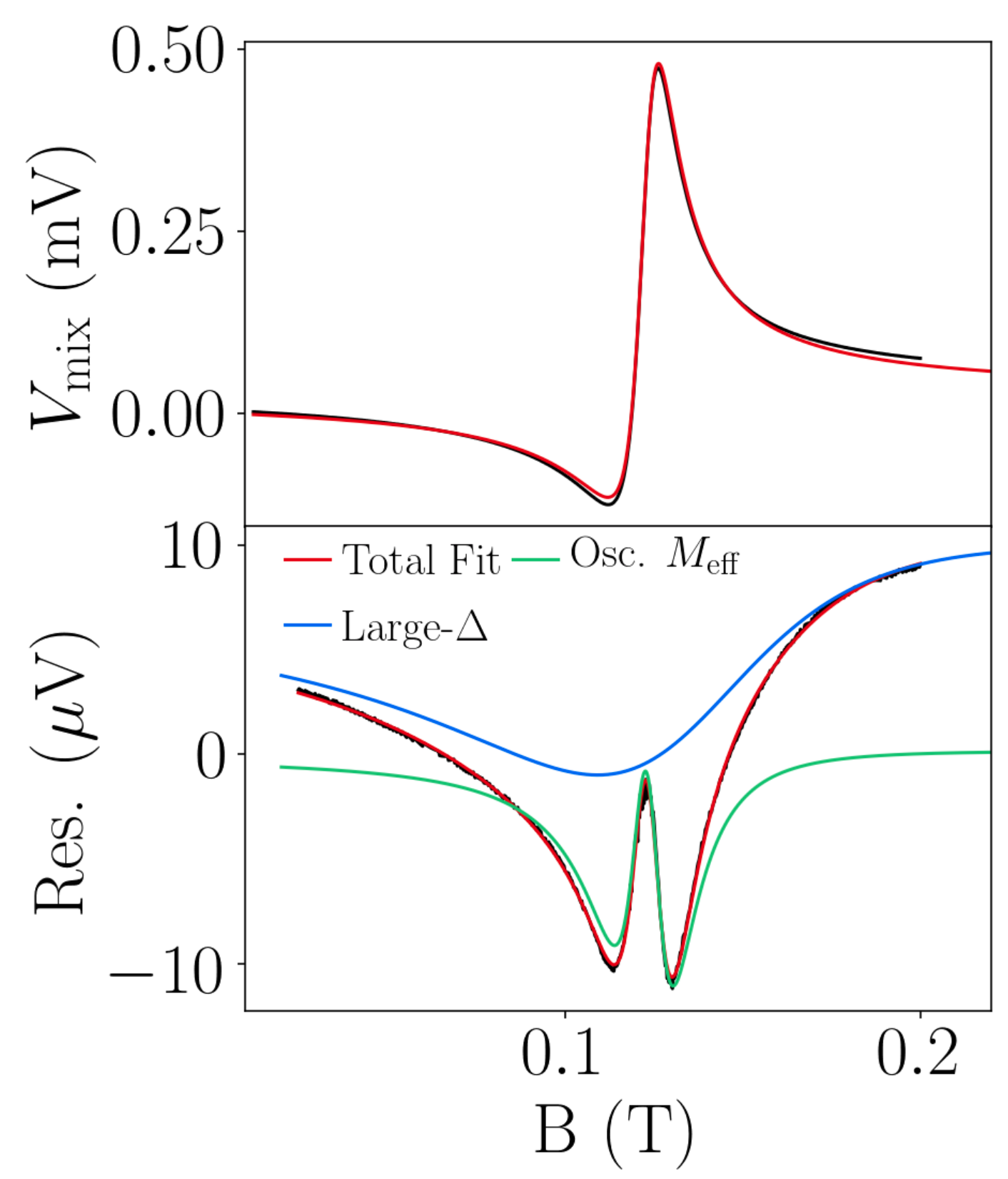}
\caption{An AM resonance with the corresponding residuals that can be fit to a fluctuating $\meff$ and large-$\Delta$ resonance. The torque ratio (S/A) in the fit to the AM resonance is held fixed to the best-fit results from the corresponding FM resonance.}
\label{delta2}
\end{figure}
\clearpage

\section{Comparisons of current-induced changes in linewidth for all P\MakeLowercase{t}/P\MakeLowercase{y} samples: both AM and FM}
\begin{figure}[h!]
\includegraphics[width=0.97\linewidth]{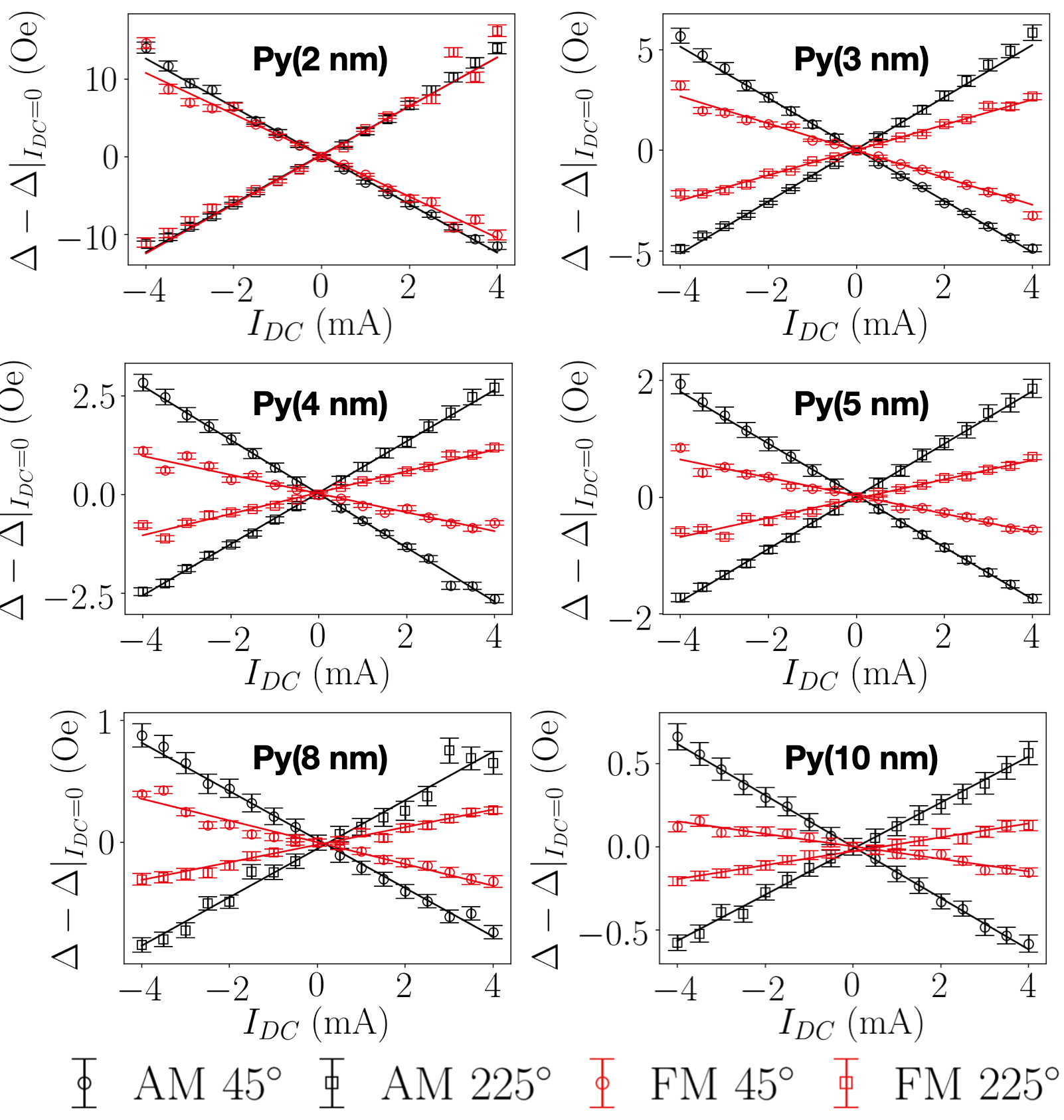}
\caption{AM and FM LW method plots for all of the samples in the Py thickness series. Each plot is equivalent to main text Fig.\ 4. The best-fit slopes of the lines in these figures are used to produce the data shown in main text Fig.\ 5.}
\label{xplots}
\end{figure}
\clearpage

\section{Current dependence of the linewidth for P\MakeLowercase{t}/C\MakeLowercase{o}F\MakeLowercase{e}B: both AM and FM} 
\begin{figure}[h!]
\includegraphics[width=0.75\linewidth]{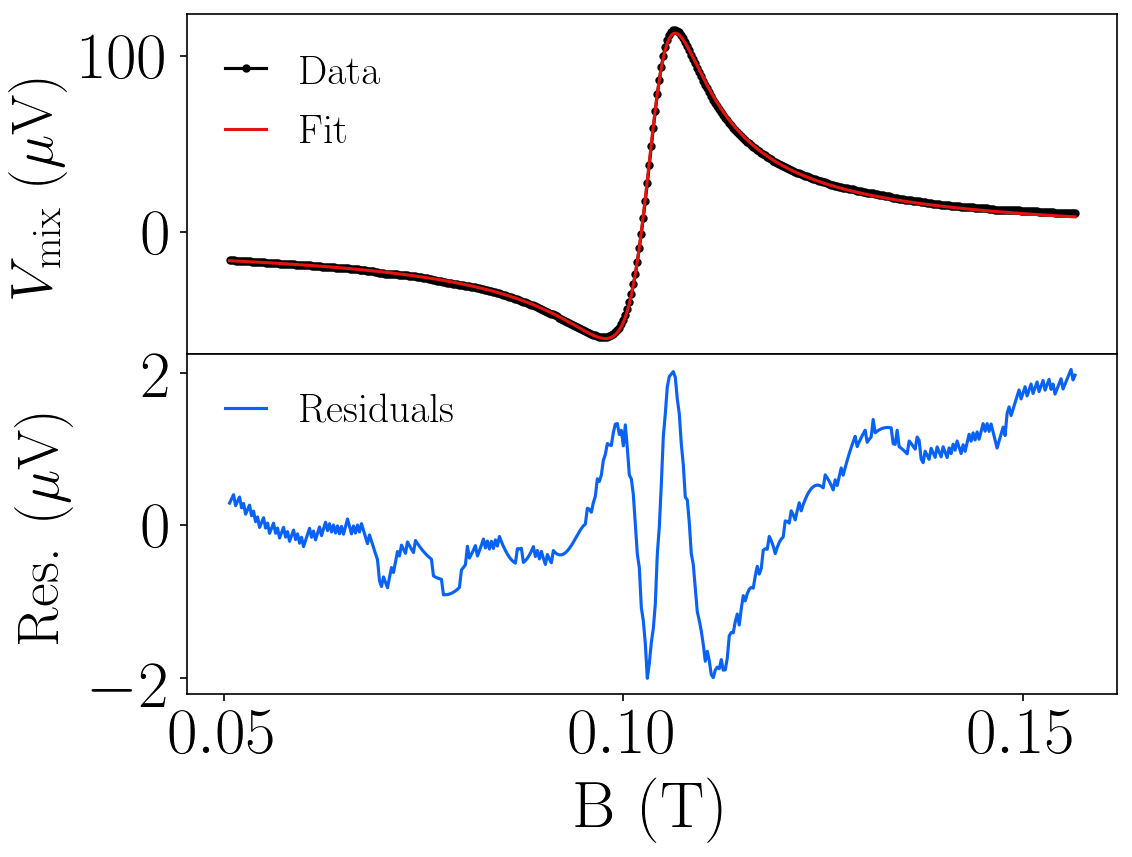}
\caption{ST-FMR measurement, best-fit, and fit residuals for a Pt(6 nm)/Co$_{40}$Fe$_{40}$B$_{20}$(6 nm) sample with a fit window of [$B_0 - 13\Delta$, $B_0+13\Delta$].}
\label{cfb0}
\end{figure}

\begin{figure}[h!]
\includegraphics[width=0.75\linewidth]{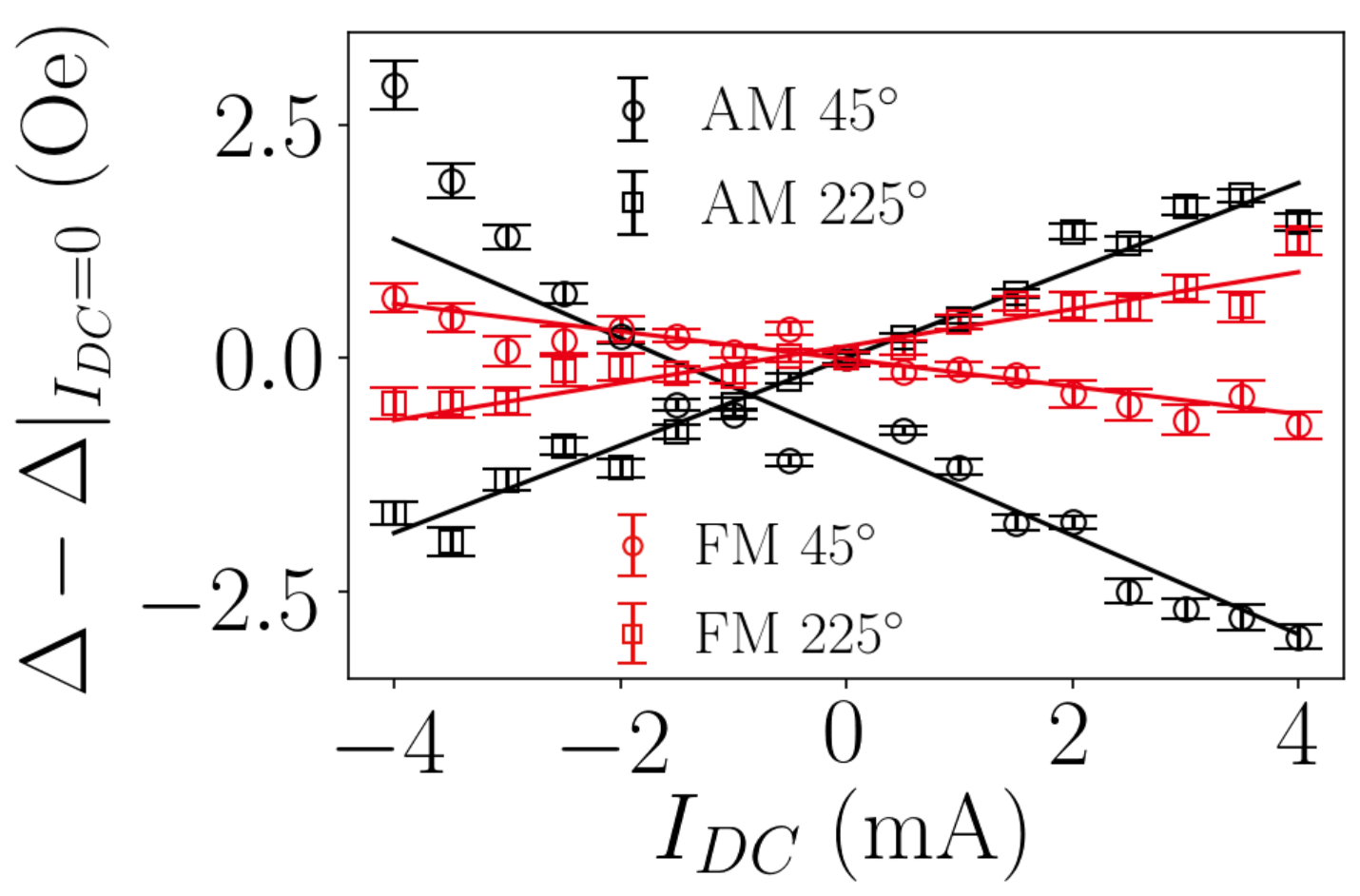}
\caption{Current dependence of the ST-FMR linewidths for a Pt(6 nm)/Co$_{40}$Fe$_{40}$B$_{20}$(6 nm) sample for a fit window of [$B_0 - 13\Delta$, $B_0+13\Delta$].  }
\label{cfb}
\end{figure}

Figure \ref{cfb0} displays the signal for an amplitude-modulated ST-FMR measurement on a Pt(6 nm)/Co$_{40}$Fe$_{40}$B$_{20}$(6 nm) sample along with the residual from a standard ST-FMR fit.
Figure \ref{cfb} shows the current dependence of the ST-FMR linewidth for the same sample, comparing the results of amplitude modulation and frequency modulations. Like the results for Pt/Py shown in the main text, the slope of the amplitude-modulated results is much larger than for the frequency-modulated measurements. We also see strong non-linearity and noise in the AM data in this sample-- making the FM LW measurement more robust.

Table \ref{cfbLW} shows the results of $\xidl$ for the Pt(6 nm)/Co$_{40}$Fe$_{40}$B$_{20}$(6 nm) sample calculated from the slopes of the lines in Fig.\ \ref{cfb}. Again like the results for Pt/Py shown in the main text, we see that the FM measurements are in much better agreement with the LS result of a Pt(6 nm)/Co$_{40}$Fe$_{40}$B$_{20}$ thickness series, as performed in ref.\  \cite{Karimeddiny2020} using the same basic procedure as the LS method in this work.

\begin{figure}[h!]
\includegraphics[width=0.75\linewidth]{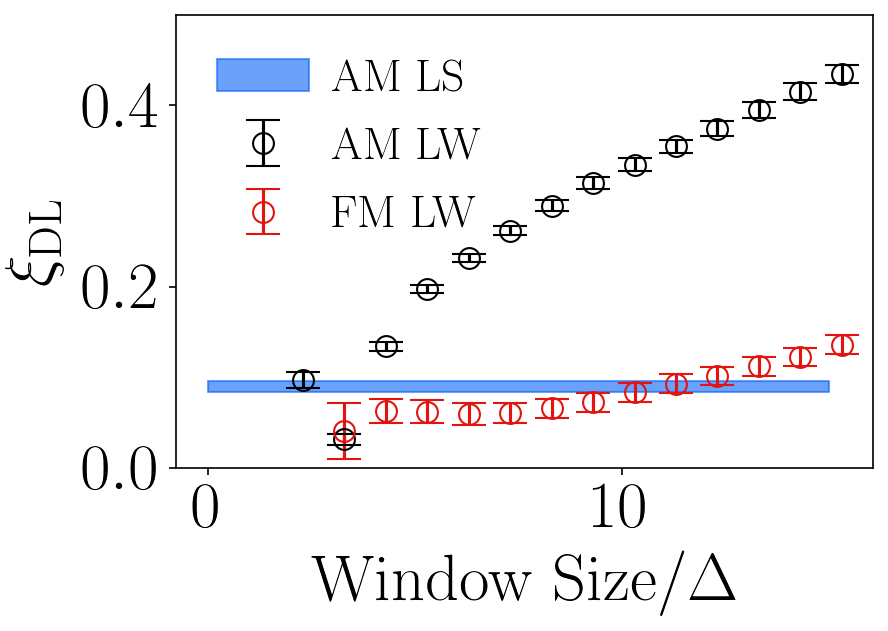}
\caption{Dependence of the results of the linewidth analysis on fit-window size for the Pt(6 nm)/Co$_{40}$Fe$_{40}$B$_{20}$(6 nm) sample. }
\label{cfbwindow}
\end{figure}

Figure \ref{cfbwindow} shows how the determination of $\xidl$ from the linewidth method for the Pt(6 nm)/Co$_{40}$Fe$_{40}$B$_{20}$(6 nm) sample  depends on the choice of fit window size used for determining the linewidths.  With a reduced fit window size, the results for both the amplitude-modulation and frequency-modulation measurements extrapolate close to the value from a lineshape analysis, in a way similar to the results for the Pt/Py samples described in the main text.

\begin{table}[h]
    \centering
    \begin{tabular}{c|cc}
        {\bf $\xidl$} & AM  & FM \\
            \hline
        LW 45$^\circ$  &  0.389(9) & 0.11(1)\\
        LW 225$^\circ$ & 0.345(7) & 0.14(1) \\
        LS (ref. \cite{Karimeddiny2020}) & 0.09(6) & --
    \end{tabular}
    \caption{Table of $\xidl$ values for a Pt(6 nm)/Co$_{40}$Fe$_{40}$B$_{20}$(6 nm) from the LW method compared with the LS method (done in ref. \cite{Karimeddiny2020}). The LW values are extracted from the slopes of the best fit lines in Fig.\ \ref{cfb}}
    \label{cfbLW}
\end{table}

\section{The Residuals of the AM mixing signal}
\subsection{$M_\text{eff}$ Oscillations}
We suggest that even a small temperature oscillation due to the AM can result in an oscillation in $M_\text{eff}$ that can be detected by the lock-in amplifier (see Main Text Fig.\ 6). If the value of $M_\text{eff}$ oscillates at the modulation frequency we have 
\begin{align}
    V_\text{mix} = V_\text{mix}\bigg\rvert_{M_\text{eff} = M^0_\text{eff}} + \frac{\partial V_\text{mix}}{\partial M_\text{eff}}\bigg\rvert_{M_\text{eff} = M^0_\text{eff}}\delta M_\text{eff}
\end{align}
with $M^0_\text{eff}$ being the equilibrium effective magnetization. We can simply differentiate the mixing signal to find
\begin{align}
    \frac{\partial V_\text{mix}}{\partial M_\text{eff}} = 
    \frac{\partial V_s}{\partial M_\text{eff}} S + 
    \frac{\partial V_a}{\partial M_\text{eff}} A + 
    \frac{1}{\Delta}\frac{\partial H_0}{\partial M_\text{eff}}
    (2V_s S A - V_a S + 2 V_a A^2).
\end{align}
In order to quantitatively estimate the magnitude of this contribution, we substitute in the fitted values of the parameters from the fits to the standard $V_\text{mix}$ signal (main text Eqs. (6) \& (7)). We find that $\frac{\partial V_\text{mix}}{\partial M_\text{eff}}\sim 10$ mV/T for the Pt(6 nm)/Py(3 nm) sample. Based on direct measurements of $M_\text{eff}$ versus temperature performed by measuring the ST-FMR resonance field while heating the sample externally (supplementary Fig.\ 12), a temperature oscillation of about 1$^{\circ}$C is required to fit this contribution to the AM residual signal

Although the oscillation $M_\text{eff}$ gives a substantial contribution to the structure of the AM residuals, we find that it contributes little to the quantitative errors in the LW AM ST-FMR analysis. The large-$\Delta$ resonance contribution to the AM residuals has the dominant effect in altering the fits to the current-dependence of the ST-FMR linewidth.
\begin{figure}[h!]
\includegraphics[width=0.7\linewidth]{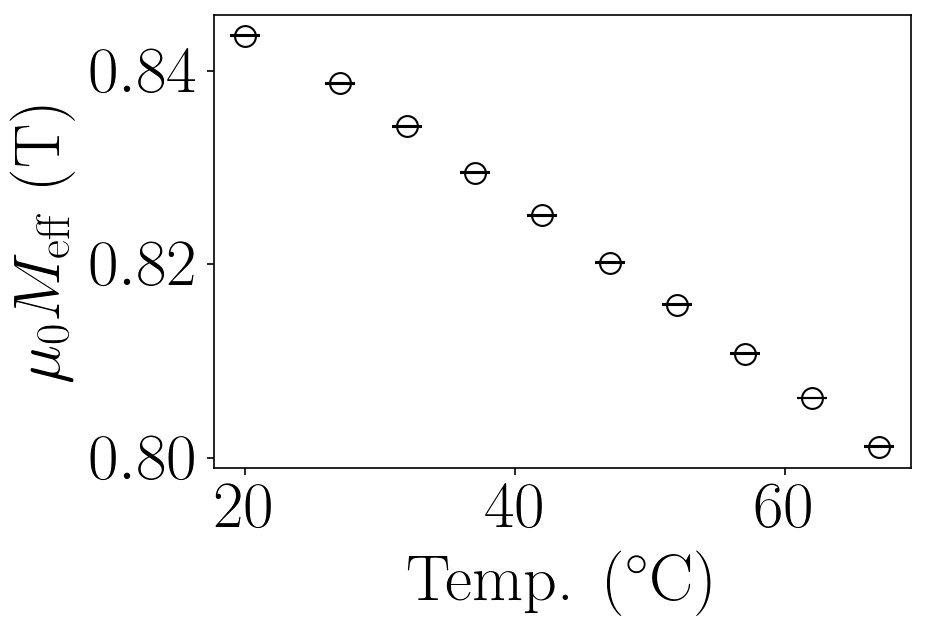}
\caption{}
\label{Meff_heat}
\end{figure}

\subsection{Large-$\Delta$ Resonance}
Determination of the parameters of the large-$\Delta$ resonance requires some care because of the potential for near-degenerate fitting parameters if one attempts to perform a combined fit to the primary resonance and the large-$\Delta$ resonance all at once.  For the following graphs, we have first determined the linewidth of the primary resonance by fitting the ST-FMR signal to a single resonance using different choices of fitting window for the magnetic field, and extrapolating the fit window to zero.  We then determine the symmetric and antisymmetric Lorentzian amplitudes for the primary resonance from a fit to the full window of data, and subtract to yield the residual signal.  The parameters of the large-$\Delta$ resonance are then determined from a fit to the residual signal that includes both an oscillating-magnetization contribution (with linewidth fixed to be the same as the primary resonance) and a large-$\Delta$ resonance signal.

\begin{figure}[h!]
\includegraphics[width=\linewidth]{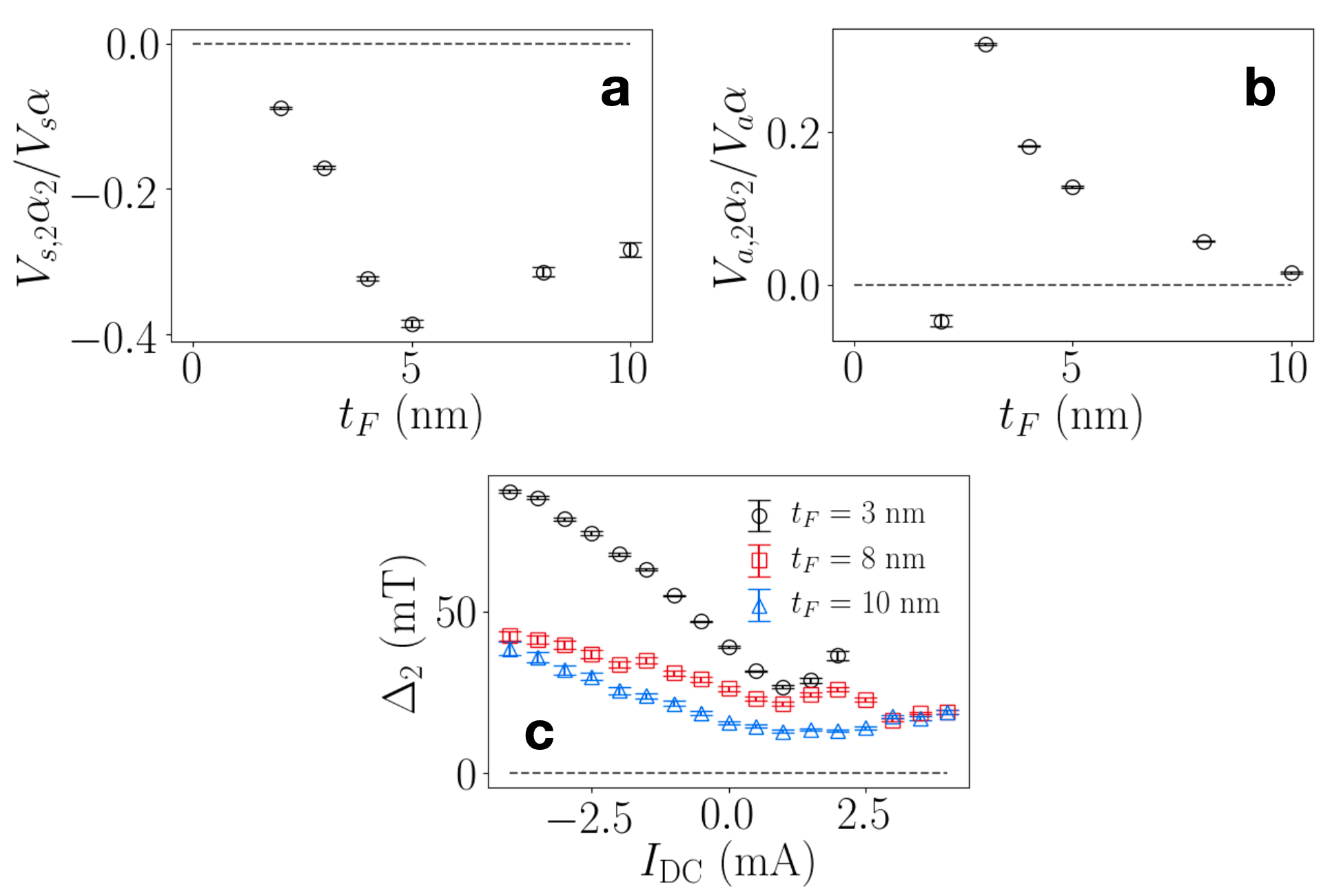}
\caption{Parameters of the large-$\Delta$ resonance that contributes to the residual in the AM fits, for the Pt/Permalloy samples. (a) Amplitude of the symmetric component of the large-$\Delta$ resonance relative to the primary resonance, normalized by the respective damping parameters, as a function of Permalloy thickness.  (b) Amplitude of the antisymmetric component of the large-$\Delta$ resonance relative to the primary resonance, normalized by the respective damping parameters, as a function of Permalloy thickness. (c) Dependence of the large-$\Delta$ linewidth on DC current for different thicknesses of the Permalloy layer.}
\label{delta2}
\end{figure}

\section{LW (in)Dependence on AM Measurement Parameters}
\begin{figure}[h]
\includegraphics[width=\linewidth]{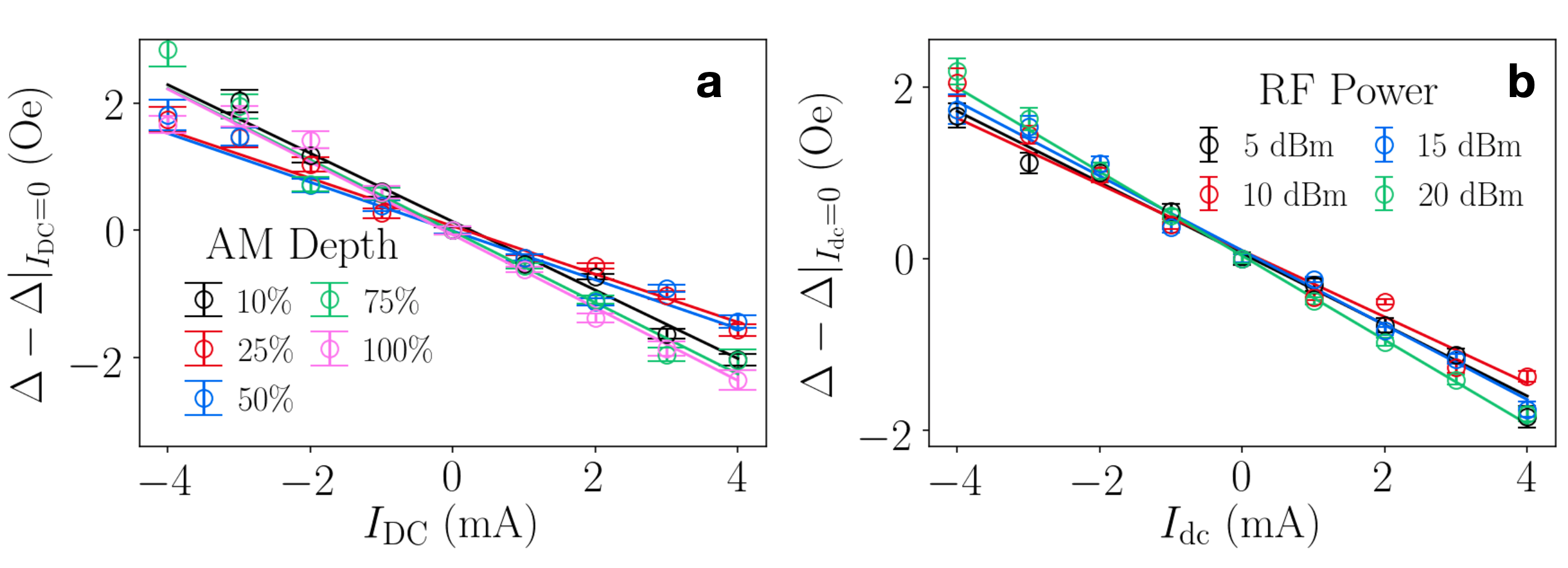}
\caption{The AM LW method performed on a Pt(6 nm)/Py(5 nm) sample (same as shown inthe main text) for {\bf (a)} a varying AM depth and {\bf (b)} a varying applied microwave power. Neither plot shows a significant systematic trend upon varying the parameter.}
\label{depthpower}
\end{figure}

In this section, we show LW plots akin to the main text Fig.\ 3, but with some variations in the parameters of the AM modulation. First, we show the dependence on the AM depth in Fig.\ \ref{depthpower}(a). All the data in the main text is taken at 100\% AM depth to maximize the SNR. An AM current may be written 
\begin{equation}
  I(t) = (1+\mu\cos\omega_m t)\sin\omega_c t .
\end{equation}
Here, the angular frequencies are as defined in the main text, and $\mu \in \left[0,1\right]$ is the AM depth, which is usually expressed as a percentage like in Fig.\ \ref{depthpower}(a).
We can see that there is very little dependence of the LW result (slope of best-fit lines) on the AM depth and that this dependence is not even monotonic, indicating that the the AM depth does not systematically affect the discrepancy observed in the AM LW measurement.

We also show dependence on the RF power applied in Fig.\ \ref{depthpower}(b). All of the main text data is taken at 20 dBm to maximize the SNR.
We can see that there is no significant dependence of the LW result on the applied RF power, which additionally confirms that our applied power is still within the linear regime.
\clearpage

\section{Dependence of LW Measurement on Fit Window for the P\MakeLowercase{t}/P\MakeLowercase{y} samples}
\begin{table}[h]
    \centering
    \begin{tabular}{c|ccccccc}
        {\bf $\xidl$} & LS & Py(2)  & Py(3) & Py(4) & Py(5) & Py(8) & Py(10)\\
            \hline
        AM 45$^\circ$  & 0.0650(3) & 0.052(2) & 0.064(1) & 0.060(1) & 0.061(1) & 0.065(1) & 0.060(1)\\
        FM 225$^\circ$ & 0.0834(7) & 0.094(6) & 0.077(2) & 0.081(1) & 0.080(4) & 0.086(4) & 0.056(3)\\
    \end{tabular}
    \caption{The results of extrapolating the windowed results shown in Main Text Fig.\ 7 to zero window size.}
    \label{cfbLW}
\end{table}

\clearpage
\newpage

 

\end{document}